%% file: paper.tex
\documentclass[preprint]{sig-alternate}

\toappear{}

\usepackage{times}
\usepackage[english]{babel}

\makeatletter
\newif\if@restonecol
\makeatother

\usepackage[ruled, vlined, linesnumbered, nofillcomment]{algorithm2e}

\usepackage{epsfig}

\usepackage{subfigure}

\usepackage{booktabs}
\usepackage{url}
\usepackage{cite}
\usepackage{xspace}
\usepackage{amsmath}
\usepackage{amssymb}
\usepackage{verbatim}
\usepackage{graphicx}
\usepackage{rotating}
\usepackage{microtype}
\usepackage{tikz}
\usetikzlibrary{snakes,arrows,shapes,positioning,fit,calc}
\usepackage{pgfplots}

\input{defines}

\newif\ifapx
\apxtrue
\usepackage[noprobabbrvs,noprobprefix,notheorems,noalgnames,noalgpseudocodecmds,draft]{pauli_all}

\title{The Long and the Short of It:\\
Summarising Event Sequences with Serial Episodes}

\numberofauthors{1}
\author{
Nikolaj Tatti \quad \quad \quad Jilles Vreeken\\
\\
\affaddr{Department of Mathematics and Computer Science}\\
\affaddr {Universiteit Antwerpen}\\
\email{\{firstname.lastname\}@ua.ac.be}
}

\date{}

\begin{document}
\maketitle

\input{abstract}

\section*{Categories and Subject Descriptors}
% -- space-compacted, fits on one line
H.2.8\,[\textbf{Database\,management}]:\,Database\,applications--\textit{Data\,mining}
% -- official, doesnt fit on one line
%\category{H.2.8}{Database management}{Database applications}[Data mining]

\terms{Theory, Algorithms, Experimentation}


\input{introduction}
\input{theory}
\input{align}
\input{search}

\input{related}
\input{experiments}
\input{discussion}
\input{conclusion}

\section*{Acknowledgements}

Nikolaj Tatti and Jilles Vreeken are supported by Post-Doctoral Fellowships of the Research Foundation -- Flanders (\textsc{fwo}).

\bibliographystyle{abbrv}
\bibliography{bib/abbrev,bib/bib-jilles,bib/bibliography}

\ifapx
\input{appendix}

\fi

\end{document}

%% file: defines.tex
\newcommand{\set}[1]{\left\{#1\right\}}

\newcommand{\enset}[2]{\left\{#1 ,\ldots , #2\right\}}

\newcommand{\define}{\leftarrow}

\newcommand{\ifam}[1]{\mathcal{#1}}

\newcommand{\CT}{\mathit{CT}}
\newcommand{\ST}{\mathit{ST}}

\newtheorem{theorem}{Theorem}

\newtheorem{proposition}[theorem]{Proposition}

\SetKw{False}{false}
\SetKw{True}{true}
\SetKw{None}{none}
\SetKw{Continue}{continue}
\SetKw{AND}{and}
\SetKw{OR}{or}
\SetKw{EACH}{each}

\SetKwInOut{Input}{input}
\SetKwInOut{Output}{output}
\SetArgSty{textnormal}

\newcommand{\given}{\mid}
\newcommand{\AB}{\Omega}
\newcommand{\DB}{D}
\renewcommand{\P}{\mathcal{P}}

\newcommand{\Model}{M}
\newcommand{\Models}{\mathcal{M}}

\newcommand{\Cands}{\mathcal{F}}

\newcommand{\Patterns}{\mathcal{P}}

\newcommand{\pieces}{\mathit{fills}}
\newcommand{\code}{\mathit{code}}
\newcommand{\usage}{\mathit{usage}}
\newcommand{\gaps}{\mathit{gaps}}
\newcommand{\DS}{\mathit{C}}

\newcommand{\LN}{L_\mathbb{N}}
\newcommand{\LP}{L_\mathit{U}}
\newcommand{\gain}{\mathit{gain}}
\newcommand{\nextw}{\mathit{next}}
\newcommand{\opt}{\mathit{opt}}
\newcommand{\diff}{\mathit{diff}}
\newcommand{\supp}{\mathit{supp}}

\newcommand{\SQS}{\textsc{Sqs}\xspace}
\newcommand{\SQSCands}{\textsc{Sqs-Candidates}\xspace}
\newcommand{\SQSSearch}{\textsc{Sqs-Search}\xspace}

\xspaceaddexceptions{$}

\makeatletter 
\pgfdeclareshape{event}{ 
\inheritsavedanchors[from=rectangle] % this is nearly a rectangle 
\inheritanchorborder[from=rectangle] 
\inheritanchor[from=rectangle]{center} 
\inheritanchor[from=rectangle]{north} 
\inheritanchor[from=rectangle]{south} 
\inheritanchor[from=rectangle]{south east} 
\inheritanchor[from=rectangle]{south west} 
\inheritanchor[from=rectangle]{west} 
\inheritanchor[from=rectangle]{east} 
\backgroundpath{% this is new 
\southwest \pgf@xa=\pgf@x \pgf@ya=\pgf@y 
\northeast \pgf@xb=\pgf@x \pgf@yb=\pgf@y 
\pgf@yc=\pgf@ya \advance\pgf@yc by 2pt 
\pgfpathmoveto{\pgfpoint{\pgf@xa}{\pgf@yc}} 
\pgfpathlineto{\pgfpoint{\pgf@xa}{\pgf@ya}} 
\pgfpathlineto{\pgfpoint{\pgf@xb}{\pgf@ya}} 
\pgfpathlineto{\pgfpoint{\pgf@xb}{\pgf@yc}} 
} 
}
\makeatother 

\newcommand{\eventseq}[2]{%
\foreach \x [count=\xi from 1, count = \xprev from 0]  in {#1}{%
    \node[draw,shape=event, right=0 of #2\xprev.south east, anchor = south west, inner sep = 2pt, outer sep = 0pt, minimum width = 11pt] (#2\xi) {\x};%
}%
}

\tikzstyle{tile} = [rounded corners = 2pt, inner sep = 0pt, fill opacity = 0.3, anchor = south west, minimum width = 11pt, minimum height = 7pt]

\newcommand{\tileseq}[2]{%
\foreach \x / \y / \z [count=\xi from 1, count = \xprev from 0]  in {#1}{%
    \node[tile, right=0 of #2\xprev.south east, anchor = south west, draw=\x, fill=\x!80, minimum width = \y, text opacity = 1, text = \x, text height=1.4ex,text depth=.5ex] (#2\xi) {$\scriptstyle \z$};%
}%
}

\newcommand{\aligntile}[5]{\node[tile, below=0.1ex of #2.south west, anchor = north west, draw=#3, fill=#3!80, minimum width = #4, text opacity = 1, text = #3, text height=1.4ex,text depth=.5ex, align = left, text width = #5]  {$\scriptstyle{#1}$};}

\newcommand{\gapseq}[2]{%
\foreach \x / \y / \z [count=\xi from 1, count = \xprev from 0]  in {#1}{%
    \node[tile, right=0 of #2\xprev.south east, anchor = south west, draw=\x, fill=\x!80, minimum width = \y,  text height=1.4ex,text depth=.5ex] (#2\xi) {};%
    \node[rectangle, right=0 of #2\xi.center, anchor = center, fill=\z, draw=\x, minimum size = 4pt, inner sep = 0]  {};%
}%
}

\pgfdeclarelayer{background}
\pgfdeclarelayer{foreground}
\pgfsetlayers{background,main,foreground}

\tikzstyle{block} = [rounded corners, draw=blue!70, fill=white, text width=3.3cm, minimum height=4em]
\tikzstyle{bgblock} = [rounded corners, draw=blue!70, thick, fill=blue!10, text width=3.3cm, minimum height=4em]
\tikzstyle{line} = [draw, -latex', thick,blue!70]

\definecolor{yafaxiscolor}{rgb}{0.3, 0.3, 0.3}

\definecolor{yafcolor1}{rgb}{0.4, 0.165, 0.553}
\definecolor{yafcolor2}{rgb}{0.949, 0.482, 0.216}
\definecolor{yafcolor3}{rgb}{0.47, 0.549, 0.306}
\definecolor{yafcolor4}{rgb}{0.925, 0.165, 0.224}
\definecolor{yafcolor5}{rgb}{0.141, 0.345, 0.643}
\definecolor{yafcolor6}{rgb}{0.965, 0.933, 0.267}
\definecolor{yafcolor7}{rgb}{0.627, 0.118, 0.165}
\definecolor{yafcolor8}{rgb}{0.878, 0.475, 0.686}

\newlength{\yafaxispad}
\newlength{\yaftlpad}
\newlength{\yaflabelpad}
\newlength{\yafaxiswidth}
\newlength{\yafticklen}
\makeatletter
\def\pgfplots@drawtickgridlines@INSTALLCLIP@onorientedsurf#1{}
\makeatother

\newcommand{\yafdrawaxis}[4]{
	\pgfplotstransformcoordinatex{#1}\let\xmincoord=\pgfmathresult 
	\pgfplotstransformcoordinatex{#2}\let\xmaxcoord=\pgfmathresult 
	\pgfplotstransformcoordinatey{#3}\let\ymincoord=\pgfmathresult 
	\pgfplotstransformcoordinatey{#4}\let\ymaxcoord=\pgfmathresult 
	\pgfsetlinewidth{\yafaxiswidth} 
	\pgfsetcolor{yafaxiscolor}
	\pgfpathmoveto{\pgfpointadd{\pgfpointadd{\pgfplotspointrelaxisxy{0}{0}}{\pgfqpointxy{\xmincoord}{0}}}{\pgfqpoint{-0.5\yafaxiswidth}{\yafaxispad}}}
	\pgfpathlineto{\pgfpointadd{\pgfpointadd{\pgfplotspointrelaxisxy{0}{0}}{\pgfqpointxy{\xmaxcoord}{0}}}{\pgfqpoint{0.5\yafaxiswidth}{\yafaxispad}}}
	\pgfpathmoveto{\pgfpointadd{\pgfpointadd{\pgfplotspointrelaxisxy{0}{0}}{\pgfqpointxy{0}{\ymincoord}}}{\pgfqpoint{\yafaxispad}{-0.5\yafaxiswidth}}}
	\pgfpathlineto{\pgfpointadd{\pgfpointadd{\pgfplotspointrelaxisxy{0}{0}}{\pgfqpointxy{0}{\ymaxcoord}}}{\pgfqpoint{\yafaxispad}{0.5\yafaxiswidth}}}
	\pgfusepath{stroke}
}

\pgfplotscreateplotcyclelist{yaf}{% 
{yafcolor1,mark options={scale=0.75},mark=o}, 
{yafcolor2,mark options={scale=0.75},mark=square},
{yafcolor3,mark options={scale=0.75},mark=triangle},
{yafcolor4,mark options={scale=0.75},mark=o},
{yafcolor5,mark options={scale=0.75},mark=o},
{yafcolor6,mark options={scale=0.75},mark=o},
{yafcolor7,mark options={scale=0.75},mark=o},
{yafcolor8,mark options={scale=0.75},mark=o}} 

\pgfplotsset{axis y line=left, axis x line=bottom,
	tick align=outside,
	tickwidth=\yafticklen,
	clip = false,
    x axis line style= {-, line width = 0pt, color=black!0},
    y axis line style= {-, line width = 0pt, color=black!0},
    x tick style= {line width = \yafaxiswidth, color=yafaxiscolor, yshift = \yafaxispad},
    y tick style= {line width = \yafaxiswidth, color=yafaxiscolor, xshift = \yafaxispad},
    x tick label style = {font=\scriptsize, yshift = \yaftlpad},
    y tick label style = {font=\scriptsize, xshift = \yaftlpad},
    every axis y label/.style = {at = {(ticklabel cs:0.5)}, rotate=90, anchor=center, font=\scriptsize, yshift = -\yaflabelpad},
    every axis x label/.style = {at = {(ticklabel cs:0.5)}, anchor=center, font=\scriptsize, yshift = \yaflabelpad},
    x tick label style = {font=\scriptsize, yshift = 1pt},
    grid = major,
    major grid style  = {dash pattern = on 1pt off 3 pt},
	every axis plot post/.append style= {line width=\yafaxiswidth} ,
	legend cell align = left,
	legend style = {inner sep = 1pt, cells = {font=\scriptsize}},
	legend image code/.code={%
		\draw[mark repeat=2,mark phase=2,#1] 
		plot coordinates { (0cm,0cm) (0.15cm,0cm) (0.3cm,0cm) };% 
	} 
}

%% file: abstract.tex
\begin{abstract}

An ideal outcome of pattern mining is a small set of informative patterns, 
containing no redundancy or noise, that identifies the key structure of the data at hand. Standard frequent pattern miners do not achieve this goal, as due to the pattern explosion typically very large numbers of highly redundant patterns are returned.

We pursue the ideal for sequential data, by employing a \textit{pattern set} mining approach---an approach where, instead of ranking patterns individually, we consider results as a whole. Pattern set mining has been successfully applied to transactional data, but has been surprisingly under studied for sequential data.

In this paper, we employ the MDL principle to identify the set of
sequential patterns that summarises the data best. In particular, we formalise how to encode sequential data using sets of serial episodes, and use the encoded length as a quality score. As search strategy, we propose two approaches:
the first algorithm selects a good pattern set from a large candidate set,
while the second is a parameter-free any-time algorithm that mines pattern sets directly from the data. Experimentation on synthetic and real data demonstrates we efficiently discover small sets of informative patterns.

\end{abstract}

%% file: introduction.tex
\section{Introduction}\label{sec:intro}

Suppose we are analysing an event sequence database, and are interested in its most important patterns. Traditionally, we would apply a frequent pattern miner, and mine all patterns that occur at least so-many times. However, due to the well-known pattern explosion we would then quickly be buried in huge amounts of highly redundant patterns, such that analysing the patterns becomes the problem, as opposed to the solution.

In this paper we therefore adopt a different approach. Instead of considering patterns individually, which is where the explosion stems from, we are after the \textit{set of patterns} that summarises the data best. Desired properties of such a summary include that it should be small, generalise the data well, and be non-redundant. To this end, we employ the Minimum Description Length principle~\cite{grunwald:07:book}, which identifies the best set of patterns as that set by which we can describe the data most succinctly. 

This approach has been shown to be highly successful for summarising transaction data~\cite{vreeken:11:krimp}, where the discovered patterns provide insight, as well as high performance in a wide range of data mining tasks, including clustering, missing value estimation, and anomaly detection~\cite{vreeken:11:krimp,vreeken:08:misval,smets:12:slim}. 
Sequence data, however, poses additional challenges compared to itemsets. For starters, the order of events is important, and we have to take gaps in patterns into account. As such, encoding the data given a cover, finding a good cover given a set of patterns, as well as finding good sets of patterns, is  much more complicated for sequences than for itemsets.

We are not the first to consider summarising sequential data. Existing methods, however, are different in that they require a single pattern to generate a full sequence~\cite{mannila:00:global}, do not consider gaps~\cite{mannila:00:global,bathoorn:06:freqpatset}, or do not punish gaps in patterns~\cite{lam:12:gokrimp}. In short, none of these methods take the full expressiveness of episodes into account.

As we identify the best model by best lossless compression, and we consider strings as data, standard compression algorithms such as Lempel-Ziv and Huffman coding are related~\cite{salomon:09:handbook}. While general purpose compressors can provide top-notch compression, they do not result interpretable models.
In our case, compression is not the goal, but a means: in order to summarise the data well, we are after those serial episodes that describe it most succinctly. 
We discuss related work in closer detail in Section~\ref{sec:related}.

In this paper, we introduce a statistically well-founded approach for succinctly {s}ummarising event se{q}uence{s}, or \SQS for short---{pronounced as `\textit{squeeze}'}. We formalise how to encode a sequence dataset given a set of episodes, and using MDL identify the best set as the set that describes the data most succinctly. To optimise this score, we give an efficient heuristic to determine which pattern best describes what part of your data. To find good sets of patterns, we introduce two heuristics: \SQSCands filters a given candidate collection, and \SQSSearch is a parameter-free any-time algorithm that efficiently mines models directly from data.

Experiments on real and synthetic data show \SQS efficiently discovers high-quality models that summarise the data well, correctly identify key patterns. The number of returned patterns stays small, up to a few hundred--- most importantly, though, the returned models do not show redundancy, and none of the patterns are polluted by frequent, yet unrelated, events.

Altogether, the long and the short of it is that \SQS mines small sets of the most important, non-redundant, serial episodes that together succinctly describe the data at hand.

%% file: theory.tex
\section{MDL for Event Sequences}\label{sec:theory}

In this section we formally introduce the problem we consider. 

\subsection{Preliminaries and Notation}\label{sec:theory:prelim}

As data type we consider \textit{event sequences}. A sequence
database $\DB$ over an event alphabet $\AB$ consists of $|\DB|$ sequences $S
\in \DB$. Every $S \in \DB$ is a sequence of $|S|$ events $e \in \AB$, i.e.  $S
\in \AB^{|S|}$. We write $S[i]$ to mean the $i$th event in $S$ and $S[i, j]$ to
mean a subsequence $S[i]\cdots S[j]$.  We denote by $||\DB||$ the sum of the
lengths of all $S_i \in \DB$, i.e. $||\DB|| = \sum_{S_i \in \DB}{|S_i|}$.
In this work, we do not explicitly consider time stamps, however we
can extend our framework to time stamped events.

The support of an event $e$ in a sequence $S$ is simply the number of
occurrences of $e$ in $S$, i.e. $\supp(e \mid S) = |\{i \in S | i = e\}|$.
The support of $e$ in a database $\DB$ is defined as $\supp(e \mid
\DB) = \sum_{S \in \DB}{\supp(e \mid S)}$. 

As patterns we consider serial episodes. A serial episode $X$ is a sequence of
events and we say that a sequence $S$ contains $X$ if there is a subsequence in
$S$ equal to $X$. Note that we are allowing gap events between the events of $X$.
A singleton pattern is a single event $e \in \AB$.

All logarithms in this paper are to base $2$, and we employ the usual convention of $0 \log 0 = 0$.

\subsection{MDL, a brief introduction}\label{sec:theory:mdl}

The Minimum Description Length principle (MDL)~\cite{grunwald:07:book} is a practical version of Kolmogorov Complexity~\cite{vitanyi:93:book}. Both embrace the slogan {\em Induction by Compression}. For MDL, this can be roughly described as follows.

Given a set of models $\Models$, the best model $\Model \in \Models$ is the one that minimises
$L(\Model) + L(\DB \mid \Model)$,
in which $L(\Model)$ is the length in bits of the description of $\Model$, and $L(\DB \mid \Model)$ is the length of the data when encoded with model $\Model$. 

This is called two-part MDL, or \textit{crude} MDL---as opposed to \textit{refined} MDL,
where model and data are encoded together~\cite{grunwald:07:book}. We use two-part MDL because we are specifically interested in the model: the patterns that give the best description. Further, although refined MDL has stronger theoretical foundations, it cannot be computed except for some special cases. Note that MDL requires the compression to be {\em lossless} in order to allow for fair comparison between different $\Model \in \Models$, and that we are only concerned with code lengths, not actual code words.

To use MDL, we have to define what our models $\Models$ are, how a $\Model \in \Models$ describes a database, and how we encode these in bits. 

\subsection{MDL for Event Sequences}

As models we consider \textit{code tables}. A code table is essentially a  look-up table, or dictionary, between patterns and associated codes. A code table has four columns, of which the first column contains patterns, the second column consists of codes for identifying these patterns, and the two right-most columns contain pattern-dependent codes for identifying gaps or the absence thereof within an embedding of a pattern. To ensure any sequence over $\AB$ can be encoded by a code table, we require that all the singleton events in the alphabet, $X \in \AB$, are included in a code table $\CT$. 

To refer to the different codes in $\CT$, we write $\code_p(X \mid \CT)$ when we refer to the code corresponding to a pattern $X$, as stored in the second column of a code table $\CT$. Similarly, we write $\code_g(X \mid \CT)$ and $\code_n(X \mid \CT)$ to resp. refer to the codes stored in the third and fourth column, which indicate whether or not the next symbol is part of a gap in the usage of pattern $X$. For readability, we do not write $\CT$ wherever clear from context.

Our next step is to explain our encoding scheme. As we will see later, there are typically very many ways of encoding a database, apart from using only singletons. Hence, for clarity, we will first explain our encoding scheme by considering how to \emph{decode} an already encoded database, and postpone finding a good encoding, or \textit{cover}, as well as how to find a good code table to Sections~\ref{sec:cover} and \ref{sec:search}.

\subsubsection*{Decoding a Database}

An encoded database consists of two code streams, $\DS_p$ and $\DS_g$, that follow from the cover $\DS$ chosen to encode the database. 
The first code stream, the \textit{pattern-stream}, denoted by $\DS_p$, is a list of $|\DS_p|$ codes, $\code_p(\cdot)$, for patterns $X \in \CT$ corresponding to those patterns chosen by `cover' algorithm. For example, $\code_p(a)\code_p(b)$ $\code_p(c)$ encodes the sequence `$\mathit{abc}$'. 

Serial episodes are not simple subsequences, however, as they allow for gaps. That is, a pattern $\mathit{de}$ specifies that event $d$, after possibly some other events, is followed by event $e$. As such, this pattern occurs both in sequence `$\mathit{de}$' as well as in sequence `$\mathit{dfe}$'. In the former there is no gap between the two events, and in the latter there is a gap of length one, in which event $f$ occurs. 

As such, only when we read the code for a singleton pattern $X$ we can directly unambiguously append $X$ to the sequence $S_k$ we are decoding---there can be no gap in a singleton pattern. When $X$ is a non-singleton pattern, on the other hand, we may only append the first symbol $x_1$ of $X$ to $S_k$; before appending event $x_2$ to $S_k$, we first need to know whether there is a gap between the two events in this usage of $X$, and if so, what event(s) occur in the gap. This is what the second code stream is for. This stream, the \textit{gap-stream}, denoted by $\DS_g$, is a list of codes from the third and fourth column of $\CT$, indicating whether gaps occur when decoding patterns. 

Given the gap-stream, we can determine whether the next event of $S_k$ may be read from the current pattern $X$, or we have to read a singleton pattern to fill the gap. Starting with an empty sequence for $S_k$, and assuming that we know its final length, we read the code for the first pattern $X$, $\code_p(X)$, from $\DS_p$, and append the first event $x_1$ of $X$ to the sequence we are decoding. If $X$ is a non-singleton pattern, we read from $\DS_g$ whether the next event is a gap-event, or not. If we read the gap-code $\code_g(X)$ from $\DS_g$ there is indeed a gap, after which we read from $\DS_p$ the $\code(Y)$ for the (singleton) event $Y \in \CT$ associated with this gap---and append it to $S_k$. We then read again from $\DS_g$ whether there is another gap, etc, until we encounter the no-gap code $\code_n(X)$ in $\DS_g$ indicating we should append the next symbol $x_j$ of $X$ to $S_k$. Whenever we are finished decoding pattern $X$, we read the code for the next pattern $X$ from $\DS_p$, until we have read as many events as $S_k$ should be long, after which we continue decoding $S_{k+1} \in \DB$ until $\DS_p$ is depleted and all $S \in \DB$ are reconstructed.

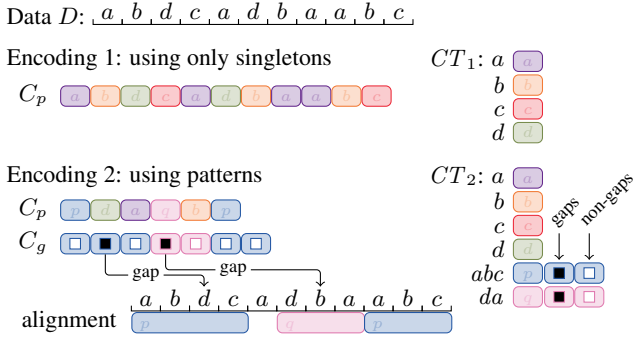
\begin{figure}[bt!]
\begin{center}
\newlength{\tilelen}
\addtolength{\tilelen}{11pt}

\begin{tikzpicture}
\node[inner sep = 0pt, minimum width = 20pt] (n0) {Data $D$:\hspace*{1mm}};
\eventseq{$a$, $b$, $d$, $c$, $a$, $d$, $b$, $a$, $a$, $b$, $c$}{n}

\node[below=9pt of n0.south west, anchor = north west, inner sep = 0pt, minimum width = 13pt] (indl) {Encoding 1: using only singletons};

\node[below=5pt of indl.south west, anchor = north west, inner sep = 0pt, minimum width = 20pt] (indp0) {$C_p$};
\tileseq{yafcolor1/\tilelen/a, yafcolor2/\tilelen/b, yafcolor3/\tilelen/d, yafcolor4/\tilelen/c, yafcolor1/\tilelen/a, yafcolor3/\tilelen/d, yafcolor2/\tilelen/b, yafcolor1/\tilelen/a, yafcolor1/\tilelen/a, yafcolor2/\tilelen/b, yafcolor4/\tilelen/c}{indp}

\node[below=22pt of indp0.south west, anchor = north west, inner sep = 0pt, minimum width = 13pt] (compl) {Encoding 2: using patterns};
\node[below=4pt of compl.south west, anchor = north west, inner sep = 0pt, minimum width = 20pt] (compp0) {$C_p$};
\tileseq{yafcolor5/\tilelen/p, yafcolor3/\tilelen/d, yafcolor1/\tilelen/a, yafcolor8/\tilelen/q, yafcolor2/\tilelen/b, yafcolor5/\tilelen/p}{compp}

\node[below=4pt of compp0.south west, anchor = north west, inner sep = 0pt, minimum width = 20pt] (compg0) {$C_g$};
\gapseq{yafcolor5/\tilelen/white, yafcolor5/\tilelen/black, yafcolor5/\tilelen/white, yafcolor8/\tilelen/black, yafcolor8/\tilelen/white, yafcolor5/\tilelen/white, yafcolor5/\tilelen/white}{compg}

\node[below=20pt of compg0.south west, anchor = north west, inner sep = 0pt, minimum width = 1.65cm] (a0) {};
\node[below=21pt of compg0.south west, anchor = north west, inner sep = 0pt, minimum width = 20pt] (al) {\hspace*{2mm}alignment};

\eventseq{$a$, $b$, $d$, $c$, $a$, $d$, $b$, $a$, $a$, $b$, $c$}{a}
\aligntile{p}{a1}{yafcolor5}{44pt}{38pt}
\aligntile{q}{a6}{yafcolor8}{33pt}{27pt}
\aligntile{p}{a9}{yafcolor5}{33pt}{27pt}

\node[above = 2.5pt of a1.north, anchor = south, inner sep = 0pt, minimum width = 13pt, minimum height = 6pt] (agap1) {\scriptsize gap};
\draw [->, rounded corners = 0.5mm, shorten >= -1pt] (agap1.east) -| (a3);
\draw [rounded corners = 0.5mm] (compg2.south) |- (agap1.west);

\node[above = 4.5pt of a4.north, anchor = south, inner sep = 0pt, minimum width = 13pt, minimum height = 6pt] (agap2) {\scriptsize gap};
\draw [->, rounded corners = 0.5mm, shorten >= -1pt] (agap2.east) -| (a7);
\draw [rounded corners = 0.5mm] (compg4.south) |- (agap2.west);

\node[right=5.5cm of indl.west, inner ysep = 1pt] (b10) {$\CT_1$: $a$};
\tileseq{yafcolor1/\tilelen/a}{b1}
\node[below=9pt of b10.south east, anchor = south east, inner ysep = 1pt] (b20) {$b$};
\tileseq{yafcolor2/\tilelen/b}{b2}
\node[below=9pt of b20.south east, anchor = south east, inner ysep = 1pt] (b30) {$c$};
\tileseq{yafcolor4/\tilelen/c}{b3}
\node[below=9pt of b30.south east, anchor = south east, inner ysep = 1pt] (b40) {$d$};
\tileseq{yafcolor3/\tilelen/d}{b4}

\node[right=5.5cm of compl.west, inner ysep = 1pt] (s10) {$\CT_2$: $a$};
\tileseq{yafcolor1/\tilelen/a}{s1}
\node[below=9pt of s10.south east, anchor = south east, inner ysep = 1pt] (s20) {$b$};
\tileseq{yafcolor2/\tilelen/b}{s2}
\node[below=9pt of s20.south east, anchor = south east, inner ysep = 1pt] (s30) {$c$};
\tileseq{yafcolor4/\tilelen/c}{s3}
\node[below=9pt of s30.south east, anchor = south east, inner ysep = 1pt] (s40) {$d$};
\tileseq{yafcolor3/\tilelen/d}{s4}
\node[below=9pt of s40.south east, anchor = south east, inner ysep = 1pt] (s50) {$\mathit{abc}$};
\tileseq{yafcolor5/\tilelen/p}{s5}
\node[right=0pt of s51.south east, anchor = south west, inner sep = 0pt, minimum size = 0pt] (g50) {};
\gapseq{yafcolor5/\tilelen/black, yafcolor5/\tilelen/white}{g5}
\node[below=9pt of s50.south east, anchor = south east, inner ysep = 1pt] (s60) {$\mathit{da}$};
\tileseq{yafcolor8/\tilelen/q}{s6}
\node[right=0pt of s61.south east, anchor = south west, inner sep = 0pt, minimum size = 0pt] (g60) {};
\gapseq{yafcolor8/\tilelen/black, yafcolor8/\tilelen/white}{g6}
\node [above=8pt, anchor = south west, rotate = 45] at (g51.north) (gl) {\scriptsize gaps};
\node [above=8pt, anchor = south west, rotate = 45] at (g52.north) (ngl) {\scriptsize non-gaps};

\draw [->, shorten >=1pt, shorten <=-3pt] (gl.south west) -- (g51);
\draw [->, shorten >=1pt, shorten <=-3pt] (ngl.south west) -- (g52);

\end{tikzpicture}\vspace*{-3pt}
\end{center}
\caption{Toy example of two possible encodings. The first encoding uses only singletons. The second encoding uses singletons and two patterns, namely, $\mathit{abc}$ and $\mathit{da}$}
\label{fig:toy}
\end{figure}

Consider the toy example given as Fig.~\ref{fig:toy}. One possible encoding would be
to use only singletons, meaning that gap stream is empty. Another encoding is
to use patterns. For example, to encode `$\mathit{abdc}$', we first give the
code for $abc$ in the pattern stream, then a no-gap code (white) in $\DS_g$ to indicate $b$, then a gap code (black) in $\DS_g$, next the code for $d$ in $\DS_p$, and we finish with a no-gap code in $\DS_g$.

\subsubsection*{Calculating Encoded Lengths}

Given the above decoding scheme we know what codes we can expect to read where and when, and hence can now formalise how to calculate the lengths of these codes, as well as the encoded lengths of the code table and database. 

We start with $L(\code_p(X))$, the lengths of pattern codes in $\DS_p$, which we can look up in the second column of $\CT$. 
By Shannon Entropy~\cite{cover:06:elements} we know that the length, in bits, of the optimal prefix-free code for an event $X$ is $-\log \Pr(X)$, where $\Pr(X)$ is the probability of $X$.
Let us write $\usage(X)$ for how often $\code_p(X)$ occurs in $\DS_p$. That is, $\usage(X) = |\{Y \in \DS_p \mid Y = \code_p(X)\}|$. 
Then, the probability of $\code_p(X)$ in $\DS_p$ is its relative occurrence in  $\DS_p$. So, we have
\[
	L(\code_p(X) \mid \CT) = -\log \left ( \frac{\usage(X)}{\sum_{Y \in \CT}{\usage(Y)}} \right ) \quad .
\]

Similarly, the lengths of the codes for indicating the presence or absence of a gap in the usage of a pattern $X$, resp. $L(\code_g(X))$ and $L(\code_n(X))$, should be dependent on their relative frequency. Let us write $\gaps(X)$ to refer to the number of gap events within the usages of a pattern $X$ in the cover of $\DB$. We then resp. have
\[
	\pieces(X) = \usage(X)(|X|-1)\quad ,
\]
for the number of non-gap codes in the usage of a pattern $X$, and
\[
	L(\code_g(X) \mid \CT) = -\log \left ( \frac{\gaps(X)}{\gaps(X) + \pieces(X)} \right ) \quad ,
\]
\[
	L(\code_n(X) \mid \CT) = -\log \left ( \frac{\pieces(X)}{\gaps(X) + \pieces(X)} \right ) \quad ,
\]
resp. for the length of a gap and a non-gap code of a pattern $X$.

We say a code table $\CT$ is code-optimal for a cover $\DS$ of a database $\DB$ if all the codes in $\CT$ are of the length according to their respective usage frequencies in $\DS_p$ and $\DS_g$ as defined above.

From the lengths of the individual codes, the encoded length of the code streams now follows straightforwardly, with resp.
\[
L(C_p \mid \CT) = \sum_{X \in \CT}{\usage(X)L(\code_p(X))}
\]
for the encoded size of the pattern-stream, and
\begin{eqnarray*}
L(C_g \mid \CT) &=& \sum_{\substack{X \in \CT \\ |X|>1}}{\Big(\gaps(X)L(\code_g(X)) + }\\[-1.5em]
&& \hphantom{\sum_{\substack{X \in \CT \\
|X|>1}}{\Big(}+} \pieces(X)L(\code_n(X))\Big)
\end{eqnarray*}
for the encoded size of the gap-stream.

Combining the above, we define $L(\DB \mid \CT)$, the encoded size of a database $\DB$ given a code table $\CT$ and a cover $\DS$, as
\begin{eqnarray*}
	L(\DB \mid \CT) &=& \LN(|\DB|) + \sum_{S \in \DB}{\LN(|S|)} + \\
	&& L(C_p \mid \CT) + L(C_g \mid \CT)\quad,
\end{eqnarray*}
where $|\DB|$ is the number of sequences in $\DB$, and $|S|$ is the length of a sequence $S \in \DB$. To encode these values, for which we have no prior knowledge, we employ the MDL optimal Universal code for integers~\cite{rissanen:83:integers,grunwald:07:book}. For this encoding, $\LN$, the number of bits required to encode an integer $n \geq 1$, is defined as 
\[
	L_{\mathbb{N}}(n) = \log^{*}(n) + \log(c_0) \quad ,
\]
where $\log^*$ is defined as $\log^*(n) = \log(n) + \log\log(n) + \cdots$, where only the positive terms are included in the sum. To make $\LN$ a valid encoding, $c_0$ is chosen as $c_0 = \sum_{J\geq1}{2^{-\LN(j)}}\approx 2.865064$ such that the Kraft inequality is satisfied.

Next we discuss how to calculate $L(\CT)$, the encoded size of a code table $\CT$. To ensure lossless compression, we need to encode the number of entries, for which we employ $\LN$ as defined above. For later use, and to avoid bias by large or small alphabets, we encode the number of singletons, $|\AB|$, and the number of non-singleton entries, $|\CT \setminus \AB|$, separately. We disregard any non-singleton pattern with $\usage(X) = 0$, as it is not used in describing the data, and has no valid (or infinite length) pattern code.

For the size of the left-hand side column, note that the simplest valid code table consists only of the single events. This code table we refer to as the \textit{standard code table}, or $\ST$. We encode the patterns in the left-hand side column using the pattern codes of $\ST$. This allows us to decode up to the names of events. 

As singletons cannot have gaps, the usage of a singleton $Y$ given $\ST$ is simply the support of $Y$ in $\DB$. Hence, the code length of $Y$ in $\ST$ is
defined as $L(\code_p(Y) \mid \ST) = -\log \frac{\supp(Y \mid \DB)}{||\DB||}$. 
Before we can use these codes, we must transmit these supports. We
transmit these using a data-to-model code~\cite{vereshchagin:03:kolmo}, an
index over a canonically ordered enumeration of all possibilities; here, the
number of ways $||\DB||$ events can be distributed over $|\AB|$ labels, where
none of the bins may be empty, as $\supp(Y \mid \DB) > 0$. The number of such
possibilities is given by ${||\DB||-1 \choose |\AB|-1}$, and by taking a log we have the number of bits required to identify the right set of
values. Note that $||\DB||$ is known from $L(\DB \mid \CT)$. In general, for
the number of bits for an index of a number composition, the number of
combinations of summing to $m$ with $n$ non-zero terms, we have 
	$\LP(m,n) = \log {m -1 \choose n-1}$,
where for $m = 0$, and $n = 0$, we define $\LP(m,n) = 0$. 

Combined, this gives us the information required to reconstruct the left-hand side of $\CT$ for the singletons, as well as the information needed to decode the non-singleton patterns of $\CT$. For a pattern $X$, the number of bits in the left-hand column is the length of $X$, $|X|$, as encoded by $\LN$, and the sum of the singleton codes
\[
\LN(|X|) + \sum_{x_i \in X}{L(\code(x_i) \mid \ST)} \quad .
\]

Next, we encode the second column. To \emph{avoid} bias, we treat the singletons and non-singleton entries of $\CT$ differently. Let us write $\P$ to refer to the non-singleton patterns in $\CT$, i.e. $\P = \CT \setminus \AB$.
For the elements of $\P$, we first encode the sum of their usages, denoted by $\usage(\P)$,
and use a data-to-model code like above to identify the correct set of individual usages. With these values, and the singleton supports we know from $\ST$, we can reconstruct the usages of the singletons in $\CT$, and hence reconstruct the pattern codes associated with each pattern in $\CT$.

This leaves us the gap-codes for the non-singleton entries of $\CT$. For reconstructing these, we need to know $\gaps(X)$, which we encode using $\LN$. The number of non-gaps then follows from the length of a pattern $X$ and its usage. As such, we can determine $\code_g(X)$ and $\code_n(X)$ exactly.

Putting this all together, we have $L(\CT \mid \DS, \DB)$, the encoded size in bits of a code table $\CT$ for a cover $\DS$ of a database $\DB$, as
\[
\begin{split}
	 L(\CT \mid \DS) = & \LN(|\AB|) + \LP(||\DB||, |\AB|) + \\
	 & \LN(|\P|+1) + \LN(\usage(\P) + 1) + \\ 
	 & \LP(\usage(\P), |\P|) + 
	 \sum_{X \in \P} L(X, \CT) \quad ,
\end{split}
\]
where $L(X, \CT)$, the number of bits for encoding the events, length, and the number of gaps of patterns $X$ in $\CT$, is
\[
\begin{split}
	&L(X, \CT)  \\
	&\quad = \LN(|X|) + \LN(\gaps(X)+1) + \sum_{x \in X}{L(\code_p(x \mid \ST))}\quad .
\end{split}
\]

By MDL, we can then define the optimal set of serial episodes for a given sequence database as the set for which the optimal cover and associated optimal  code table minimises the total encoded size 
\[
L(\CT,\DB) = L(\CT \given \DS) + L(\DB \given \CT) \; .
\]

More formally, we define the problem as follows.

\vspace{1em}\textbf{Minimal Code Table Problem}
{\em Let $\AB$ be a set of events and let $\DB$ be a sequence database over $\AB$, 
find the minimal set of serial episodes $\P$ such that for the optimal cover $\DS$ of $\DB$ using $\P$ and $\AB$, the total encoded cost $L(\CT, \DB)$ is minimal, 
where $\CT$ is the code-optimal code table for $\DS$.
}\vspace{1em}

Clearly, this problem entails a rather large search space. First of all, given
a set of patterns, there are many different ways to cover a database. Second,
there are very many sets of serial episodes $\P$ we can consider, namely all
possible subsets of the collection of serial episodes that occur in
$\DB$. However, neither the full problem, or these  sub-problems, exhibit
trivial structure that we can exploit for fast search, e.g. (weak) monotonicity. 

We hence break the \textbf{Minimal Code Table Problem} into two sub-problems. First, in the next section we discuss how to optimise the cover of a sequence \textit{given} a set of episodes. Then, in Section~\ref{sec:search}, we will discuss how to mine high quality code tables.

%% file: align.tex
\section{Covering a string}\label{sec:cover}

Encoding, or covering, a sequence is more difficult than decoding one. The reason is simple: when decoding there is no ambiguity, while when encoding there are many choices, i.e. what pattern to encode a symbol with. In other words, given a set of episodes, there are many valid ways to cover a sequence, where by our problem definition we are after the cover $\DS$ that minimises $L(\CT, \DB)$.

Due to lack of space, we provide the proofs in 
\ifapx 
Appendix~\ref{sec:apx}.
\else 
the Appendix\footnote{\url{http://adrem.ua.ac.be/sqs/}}.
\fi

\subsection{Minimal windows}

% alignment
Assume we are decoding a sequence $S_k \in \DB $. Assume we decode the beginning of a pattern $X$ at $S_k[i]$ and that the last symbol belonging to this instance of $X$ is, say, $S_k[j]$. We say that $S_k[i, j]$ is an \emph{active} window for $X$. 
Let $\P$ be the set of non-singleton patterns used by the encoding.
We define an \emph{alignment} $A$ to be the set of all active windows for all non-singleton patterns $X \in \P$ as
\[
	A = \set{(i, j, X, k) \mid S_k[i, j] \text{ is an active window for } X, 
S_k \in D} \; .
\]
An alignment corresponding to the second encoding given in Figure~\ref{fig:toy} is
$\{(1, 4, \mathit{abc}, 1), (6, 8, \mathit{da}, 1), (9, 11, \mathit{abc}, 1)\}$.

Note that an alignment $A$ does not uniquely define the cover of the
sequence, as it does not take into account how the intermediate symbols (if
any) within the active windows of a pattern $X$ are encoded. However, an alignment $A$ for a sequence database $\DB$ does define an equivalence class over covers of the same encoded length. In fact, given a
sequence database $\DB$ and an alignment $A$, we can determine the number of
bits our encoding scheme would require for such a cover. To see this, let $X$ be a pattern and let $W = \set{(i, j, X, k) \in A}$, then 
\begin{equation}
\label{eq:usage}
	\usage(X) = \abs{W} \text{ and } \gaps(X) = \gaps(W)\quad,
\end{equation}
where
\begin{equation}
\label{eq:gaps}
	\gaps(W) = \sum_{(i, j, X, k) \in W} j - i - (\abs{X} - 1) \quad.
\end{equation}
The remaining symbols are encoded as singleton patterns. Hence, the
usage of a singleton is equal to
\begin{equation}
\label{eq:usagesingleton}
	\usage(s) = \supp(s \mid \DB) - \sum_{s \in X} \usage(X)\quad.
\end{equation}

Given an alignment $A$ for $\DB$, we can trivially construct a valid cover $\DS$ for $\DB$, simply by following $A$ and greedily covering $S_k$ with pattern symbols if possible, and singletons otherwise. That is, if for a symbol $S_k[i]$ we have, by $A$, the choice for covering it as a gap or non-gap of a pattern $X$, we choose non-gap. 

Then, from either $\DS$, or directly from $A$, we can derive the associated code-optimal code table $\CT$.
Given an alignment $A$, let us write $\CT(A)$ for this code table.
Wherever clear from context, we will write
$L(\DB \mid A)$ to mean $L(\DB \mid \CT(A))$, and similarly
$L(\DB, A)$ as shorthand for $L(\DB,  CT(A))$.

% minimal window
Our next step is to show what kind of windows can occur in the optimal alignment.  We say
that $W = S[i, j]$ is a minimal window of a pattern $X$ if $W$ contains $X$ but
no other proper sub-windows of $W$ contain $X$. For example, in Figure~\ref{fig:toy}
$S[6,8]$ is a minimal window for $\mathit{da}$ but $S[6,9]$ is not.

% prop: active windows are minimal windows
\begin{proposition}
\label{prop:minimal}
Let $A$ be an alignment producing an optimal encoded length. Then all
active windows in $A$ are minimal windows.
\end{proposition}

% minimal windows are limited amount
Proposition~\ref{prop:minimal} says that we need to only study minimal windows.  Let
$\ifam{F}$ be a set of episodes and let $X \in \ifam{F}$. Since an event $S_k[i]$ can be a starting point to only one
minimal window of $X$, there are only $\norm{\DB}$ minimal windows of $X$ in $D$,
at most. Consequently, the number of minimal windows we need to investigate
is bounded by $\norm{\DB}\abs{\ifam{F}}$.  Moreover, we can use \textsc{FindWindows} 
in~\cite{tatti:11:clsepdami} to discover all the minimal windows for a pattern $X$ in
$O(\abs{X}\norm{\DB})$ time.

% complexity results

\subsection{Finding optimal alignment}
Discovering an optimal alignment is non-trivial due to the complex relation between code lengths and the alignment. However, if we fix the alignment,
Eqs.~\ref{eq:usage}--\ref{eq:usagesingleton} give us the codes optimising
$L(\DB \mid A)$. In this section we will show the converse, that if we fix the
codes, we can easily find the alignment optimising $L(\DB \mid A)$. In
order to do that let $w = (i, j, X, k)$ be a minimal window for a pattern $X$.
We define the gain to be

% gain of a window
\[
\begin{split}
	\gain(w) = & -L(\code_p(X)) - (j - i - \abs{X})L(\code_g(X)) \\
	&  - (\abs{X} - 1)L(\code_n(X)) + \sum_{x \in X} L(\code_p(x))\quad. \\
\end{split}
\]

% proposition: optimal alignment = disjoint windows with highest gain
\begin{proposition}
\label{prop:align}
Let $\DB$ be a dataset and $A$ be an alignment.
Then the length of encoding $D$ is equal to
\[
	L(\DB \mid A)  = \mathit{const} - \sum_{w \in A}\gain(w)\quad,
\]
where $\mathit{const}$ does not depend on $A$.
\end{proposition}

This proposition suggests that if we fix the code lengths we need to maximise the
total gain. In order for an alignment to be valid, the windows must be disjoint.
Hence, given a set of $W$, consisting of all minimal windows of the given
patterns, we need to find a subset $O \subseteq W$ of disjoint windows
maximising the gain.

Assume that $W$ is ordered by the first index of each window.  For a window
$w$, define $\nextw(w)$ to be the next disjoint window in $W$.  Let $o(w)$ be
the optimal total gain of $w$ and its subsequent windows. Let $v$ be the next window of $w$,
then the optimal total gain is $o(w) = \min(o(v), \gain(w) + o(\nextw(w))$.
This gives us a simple dynamic program, \textsc{Align}, given as Algorithm~\ref{alg:align}.

\begin{algorithm}[htb!]
\caption{\textsc{Align}$(W)$}
\label{alg:align}
\Input{minimal windows $W$ sorted by the first event}
\Output{mutually disjoint subset of $W$ having the optimal gain}

$o({N + 1}) \define 0$; $\opt({N + 1}) \define \None$\;

\ForEach {$i = N, \ldots, 1$} {
	$c \define 0$\;
	\lIf {$\nextw(i)$} {$c \define  o(\nextw(i))$ }
	\uIf {$\gain(w_i) + c > o({i + 1})$ } {
		$o(i) \define \gain(w_i) + c$; $\opt(i) \define i$\;
	}
	\Else {
		$o(i) \define o({i + 1})$; $\opt(i) \define \opt({i + 1})$\;
	}
}

$O \define $ optimal alignment (obtained by iterating $\opt$ and $\nextw$)\;

\Return {$O$\;}

\end{algorithm}

We can now use \textsc{Align} iteratively. Given the codes we find the optimal
alignment and derive the optimal codes given the new alignment. We repeat this
until convergence, which gives us a heuristic approximation to the optimal alignment $A^*$ for $\DB$ using patterns $\Patterns$. As initial values, we use the number of minimal windows as usage and set gap code length to be $1$ bit. 
The pseudo code of \SQS, which stands for Summarising event seQuenceS, is given as Algorithm~\ref{alg:sqs}.

% seqs
\begin{algorithm}
\caption{\SQS$(\DB, \Patterns)$. Summarising event seQuenceS}
\label{alg:sqs}
\Input{Database of sequences $\DB$, set of patterns $\Patterns$}
\Output{Alignment $A$}

\lForEach {$s \in \AB$} {
	$\usage(s) \define \supp(s \mid D)$
}
\ForEach {$X \in \Patterns, \abs{X} > 1$} {
	$W_X \define$ \textsc{FindWindows}$(X, \DB)$\;
	$\usage(X) \define \abs{W_X}$; $\gaps(X) \define \abs{X} - 1$\;
}
$W \define $ merge sort $\set{W_X}_{X \in \ifam{F}}$ based on first event\;

\While {changes} {
	compute gain for each $w \in W$\;
	$A \define \textsc{Align}(W)$\;
	recompute usage and gaps from $A$ (Eqs.~\ref{eq:usage}--\ref{eq:usagesingleton})\;
}
\Return {$A$\;}
\end{algorithm}

% complexity results
% finite number of iterations
% in practice small
The computational complexity of single iteration comes down to the
computational complexity of \textsc{Align}$(W)$, which is $O(\abs{W}) \subseteq
O(\abs{\Patterns}\norm{\DB})$.  Also note that $\nextw$ is precomputed before calling
\textsc{Align} and this can be also computed with a single scan, taking
$O(\abs{W})$ steps. Note that the encoded length improves at every iteration,
and as there are only finite number of alignments, \SQS will converge to a local optimum in finite time. In practice, the number of iterations is small---in the experiments typically less than $10$.

%% file: search.tex
\section{Mining Code Tables}\label{sec:search}

With the above, we both know how to score the quality of a pattern set, as well as how to heuristically optimise the alignment of a pattern set. This leaves us with the problem of finding good sets of patterns. In this section we give two algorithms to do so.

\subsection{Filtering Candidates}

Our first algorithm, \SQSCands, assumes that we have a (large) set of candidate patterns $\Cands$. In practice, we assume the user obtains this set of patterns using a frequent pattern miner, although any set of patterns over $\AB$ will do.
From this set $\Cands$ we then select that set of patterns $\Patterns \subseteq \Cands$ such that the optimal alignment $A$ and associated code table $\CT$ minimises $L(\DB, \CT)$.

For notational brevity, we simply write $L(\DB, \Patterns)$ as shorthand for the total encoded size $L(\DB, \CT)$ obtained by the code table $\CT$ containing a set of patterns $\Patterns$ and singletons $\AB$, and being code-optimal to the alignment $A$ as found by \SQS.

We begin by sorting the candidates $X \in \Cands$ by $L(\DB, \set{X})$ from
lowest to highest. 
After sorting, we iteratively greedily test each pattern $X \in \Cands$. If adding $X$ to $\Patterns$ improves the score, i.e. fewer bits are needed, we keep $X$ in $\Patterns$, otherwise it is permanently removed. The pseudo-code for \SQSCands is given as Algorithm~\ref{alg:sqscand}.

\begin{algorithm}[htb!]
\caption{\SQSCands$(\Cands, \DB)$}
\label{alg:sqscand}
\Input{candidate patterns $\Cands$}
\Output{set of non-singleton patterns $\Patterns$ that heuristically minimise the \textbf{Minimal Code Table Problem}}

order patterns $X \in \Cands$ based on $L(\DB, \set{X})$\;

$\Patterns \define \emptyset$\;
\ForEach{$X \in \Cands$ in order} {
	\If {$L(\DB, \Patterns \cup X) < L(\DB, \Patterns)$} {
		$\Patterns \define$ \textsc{Prune}$(\Patterns \cup X, \DB, \False)$\;
	}
}

$\Patterns \define$ \textsc{Prune}$(\Patterns, \DB, \True)$\;
order patterns $X \in G$ by $L(\DB, \Patterns) - L(\DB, \Patterns \setminus X)$\;
\Return {$\Patterns$}\;
\end{algorithm}

During the search we iteratively update the code table Hence, it may be that over time,
previously included patterns start to harm compression once their role in
covering the sequence is taken over by new, more specific, patterns. As such,
they become redundant, and should be removed from $\Patterns$.

To this end, we prune redundant patterns (see
Algorithm~\ref{alg:purge}) after each successful addition. During pruning, we iteratively consider each pattern $Y \in \Patterns$ in order of insertion. If $\Patterns \setminus X$ improves the total encoded size, we remove $X$ from $\Patterns$. 
As testing every pattern in $\Patterns$ at every successful addition may become rather time-consuming, we use a simple heuristic: if the total gain of the windows of
$X$ is higher than the cost of $X$ in the code table we do not test $X$.

\begin{algorithm}
\caption{\textsc{Prune}$(\Patterns, \DB, \mathit{full})$}
\label{alg:purge}
\Input{pattern set $\Patterns$, database $\DB$, boolean variable $\mathit{full}$, \\ \False for heuristic scan, \True for complete scan}
\Output{pruned pattern set $\Patterns$\;}

\ForEach{$X \in \Patterns$} {
	$\CT \define$ code table corresponding to \SQS$(\DB, \Cands)$\;
	$\CT' \define$ code table obtained from $\CT$ by deleting $X$\;
	$g \define \sum_{w = (i, j, X, k) \in A} \gain(w)$\;

	\If {$\mathit{full}$ {\OR} $g < L(\CT) - L(\CT')$} {
		\lIf {$L(\DB, \Patterns \setminus X) < L(\DB, \Patterns)$} {
			$\Patterns \define \Patterns \setminus X$
		}
	}
}
\Return {$\Patterns$}\;

\end{algorithm}

After \SQSCands considered every pattern of $\Cands$, we run one final round of
pruning without this heuristic. Finally, we order the patterns in $\Patterns$ by $L(\DB, \Patterns) - L(\DB, \Patterns\setminus X)$. That is, by the impact on the total encoded length when removing $X$ from $\Patterns$. This order tells us which patterns in $\Patterns$ are most important.

Let us consider the execution time needed by \SQSCands. Ordering patterns can be done
in $O(\abs{\Cands}\norm{D})$ time. Computing $L(\DB, \Patterns \cup X)$
can be done in $O(\abs{\Patterns}\norm{D}) \subseteq O(\abs{\Cands}\norm{D})$ time.
Pruning can be done in $O(\abs{\Patterns}^2\norm{D}) \subseteq O(\abs{\Cands}^2\norm{D})$.
Combined, this gives us a total time complexity of $O(\abs{\Cands}^3\norm{D})$.
In practice, the algorithm is much faster, however, as first, due to MDL the code tables remain small, and hence $\abs{\Patterns} \ll \abs{\Cands}$,
second, the execution time of \SQS is typically faster than $O(\abs{\Patterns}\norm{D})$,
and third, the pruning heuristic further reduces the computational burden.

\subsection{Directly Mining Good Code Tables}

The \SQSCands algorithm requires a collection of candidate patterns to be materialised, which in practice can be troublesome; the well-known pattern explosion may prevent patterns to be mined at as low thresholds as desired.
In this section we propose an alternative strategy for discovering good code tables directly from data. Instead of filtering a pre-mined candidate set, we now discover candidates on the fly, considering only patterns that we expect to optimise the score given the current alignment.

To illustrate the general idea, consider that we have a current set of patterns
$\Patterns$.  We iteratively find patterns of form $XY$, where $X, Y \in \Patterns \cup \AB$ producing the lowest $L(\DB, \Patterns \cup \set{XY})$. We add $XY$ to $\Patterns$ and continue until no gain is possible. Unfortunately, as testing each combination takes $O((\abs{\Patterns} + \abs{\AB})^2(\abs{\Patterns} + 1)\norm{D})$ time, we cannot do this exhaustively and exactly within reasonable time.

Hence, we resort to heuristics.

To guarantee the fast discovery of good candidates, we design a heuristic
algorithm that, given a pattern $P$, will find a pattern $PQ$ of high expected gain in only $O(\abs{\Patterns} + \abs{\AB} + \norm{D})$ time.

Our first step is to demonstrate that if we take $N$ active windows of $P$,
and $N$ active windows of $Q$, and convert them into $N$ active windows of $PQ$,
the difference in total encoded length can be computed in constant time. 

\begin{proposition}
\label{prop:gainconstant}
Fix a database $\DB$ and an alignment $A$. Let $P$ and $Q$ be two
patterns. Let $V = \enset{v_1}{v_N}$ and $W = \enset{w_1}{w_N}$ be two set of
candidate windows for $P$ and $Q$, respectively. Assume that either $P$ ($Q$)
is a singleton or each $v_i$ ($w_i$) occurs in $A$. Assume that $v_i$ and $w_i$
occur in the same sequence and write $v_i = (a_i, b_i, P, k_i)$ and $w_i =
(c_i, d_i, Q, k_i)$. Assume that $b_i < c_i$. Write $R = PQ$ and let
\[
	U = \enset{(a_1, d_1, R, k_1)}{(a_N, d_N, R, k_N)}\quad.
\]
Assume that $U$ has no overlapping windows and has no overlapping windows 
with $A \setminus (V \cup W)$. Then the difference
\[
	L(\DB, A \cup U \setminus (V \cup W)) - L(\DB, A)
\]
depends only $N$, $\gaps(V)$, $\gaps(W)$, and $\gaps(U)$ and can be computed
in constant time from these values.
\end{proposition}

The conditions given in Proposition~\ref{prop:gainconstant} are needed so that $A \setminus (V \cup W)$
is a proper alignment.
We denote the aforementioned difference by 
$\diff(V, W, U; A, D)$.
Note that this difference partly depends on $A$ and $\DB$. However, since we keep these fixed in the proposition they only contribute constant terms. Further note that $U$ should not overlap with  $A \setminus (V \cup W)$. We will address this limitation later. 
We should point out that in practice we do not keep lists of $U$, $V$, and $W$, but instead exploit the gap counts and number of windows, as this is sufficient for computing the difference.

Now that we have a way of computing the gain of using windows for $PQ$, we need
to know which windows to use in the alignment. The following proposition suggest that we should pick the windows with the shortest length.

\begin{proposition}
\label{prop:smallest}
Let $\DB$ be a database and $A$ be an alignment. Let $v = (i, j, X, k) \in A$.
Assume that there exists a window $S_l[a, b]$ containing $X$ such that
$w = (a, b, X, l)$ does not overlap with any window in $A$ and $b - a < j - i$.
Then $A$ is not an optimal alignment.
\end{proposition}

This proposition gives us an outline of the heuristic. We start enumerating
minimal windows of $PQ$ from shortest to largest.  At each step we compute the
score using Proposition~\ref{prop:gainconstant}, and among these scores we the
pick optimal one. 

We cannot guarantee linear time if we consider each $Q$ individu\-ally. Instead,
we scan for all candidates simultaneously. In addition, to
guarantee linear time we consider only active windows of $P$ and $Q$, and do
not consider singletons occurring in the gaps. The scan starts by finding all
the active windows (ignoring singletons in gaps) of $P$. We then continue by scanning the
patterns occurring after each $P$. We interleave the scans in such a way that
the new minimal windows are ordered, from shortest to longest. We stop the
scan after we find next occurrence of $P$ or the end of the sequence.

There are two constraints that we need to take into account. When enumerating
minimal windows of $\mathit{PQ}$ we need to make sure that we can add them to the alignment. That is, a new minimal window cannot intersect with other new
minimal windows, and the only windows it may intersect in the alignment are the two windows from which it was constructed. The first constraint can only happen
when $Q = P$, in which case we simply check if the adjacent scans have already
used these two instances of $P$ for creating a minimal window for pattern $\mathit{PP}$.
To guarantee the second constraint, we need to delete the intersecting windows
from the alignment. We estimate the effect of deleting $w$ by adding $\gain(w)$
(computed from the current alignment) to the score.
The pseudo-code for calculating this estimate is given as Algorithm~\ref{alg:estimate}.

\begin{algorithm}[htb!]
\caption{\textsc{Estimate}$(P, A, \DB)$. Heuristic for finding a pattern $X$ used by the current encoding  with a low $L(\DB, A \cup \mathit{PX})$}
\label{alg:estimate}
\Input{database $\DB$, current alignment $A$, pattern $P \in \CT$}
\Output{pattern $PX$ with $X \in CT$ and a low $L(\DB, A \cup \mathit{PX})$}

\lForEach{$X \in \CT$} {
	$\!V_X \define \emptyset$;
	$W_X \define \emptyset$;
	$U_X \define \emptyset$;
	$d_X \define 0$
}

$T \define \emptyset$\;
\ForEach {occurrence $v$ of $P$ in the encoding (ignoring gaps)} {
	$(a, b, P, k) \define v$\;
	$d \define $ the end index of the active window following $v$\;
	$\mathit{t} \define (v, d, 0)$; $l(t) \define d - a$\;
	add $t$ into $T$\;
}

\While {$T$ is not empty} {

	$t \define \arg\min_{u \in T} l(u)$\;

	$(v, d, s) \define t$; $a \define $ first index of $v$\;

	$w = (c, d, X, k) \define$ active window of a pattern ending at $d$\;
	\If {$X = P$ \AND (event at $a$ or $d$ is marked)} {
		delete $t$ from $T$\;
		\Continue\;
	}

	\If {$S_k[a, d]$ is a minimal window of $\mathit{PX}$} {
		add $v$ into $V_X$\;
		add $w$ into $W_X$\;
		add $(a, d, \mathit{PX}, k)$ into $U_X$\;

		$d_X \define \min(\diff(V, W, U; A) + s, d_X)$\;
		\lIf {$\abs{X} > 1$} { $s \define s + \gain(w)$}
		\If {$X = P$} {
			mark the events at $a$ and $d$\;
			delete $t$ from $T$\;
			\Continue\;
		}
	}
	\If {$w$ is the last window in the sequence} {
		delete $t$ from $T$\;
	}
	\Else {
		$d \define $ the end index of the active window following $w$\;
		update $t$ to $(v, d, s)$ and $l(t)$ to $d - a$\;
	}
}
\Return {$\mathit{PX}$ with the lowest $d_X$\;}
\end{algorithm}

\begin{proposition}
\label{prop:optimal}
\textsc{Estimate}$(P, \emptyset, D)$ returns a pattern with optimal score. 
\end{proposition}

Next, let us consider the computational complexity of this approach. The initialisation in
\textsc{Estimate} can be done in $O(\abs{\AB} + \abs{\Patterns} + \norm{D})$,
where $\Patterns$ are the current non-singleton patterns. After selecting
the next window, each step in the main loop can be done in constant time.
The only non-trivial step is picking the next smallest window. However, since
the window lengths are integers smaller or equal than $\norm{D}$, we can store the candidates into an array of lists, say $N_d$, where $N_d$ contains the windows of length $d$. Finding the next window may take more than a constant time since
we need to find the next non-empty list $N_d$ but such search may only
contribute $\norm{D}$ checks in total. Since we stop after encountering $P$, every
event is visited only twice at maximum, hence the running time for \textsc{Estimate}
is $O(\abs{\AB} + \abs{\Patterns} + \norm{D})$.

The actual search algorithm, \SQSSearch, calls \textsc{Estimate} for each
pattern $P$. The algorithm, given as Algorithm~\ref{alg:search}, continues by
sorting the obtained patterns based on their estimated scores and attempts to
add them into encoding in the same fashion as in \SQSCands. After each
successful addition of pattern $X$, we scan for the gap events occurring in the
active windows of $X$, and test patterns obtained from $X$ by adding a gap
event as intermediate event. The scan can be done in $O(\norm{D})$ time, and in
theory we may end up testing $\abs{\AB}(\abs{X} - 1)$ patterns. In practice,
the number is much smaller since accepted patterns typically have small gaps.
If any of these patterns in successfully added we repeat this procedure in a
recursive fashion.
In practice, testing $X$ is relatively fast, and the
total computational complexity is dominated by \textsc{Estimate}.

\begin{algorithm}[htb!]
\caption{\SQSSearch$(\DB)$}
\label{alg:search}
\Input{database $\DB$}
\Output{significant patterns $\Patterns$}

$\Patterns \define \emptyset$;
$A \define \SQS(\DB, \emptyset)$\;

\While {changes} {
	$\Cands \define \emptyset$\;
	\lForEach {$P \in \CT$} {
		add \textsc{Estimate}$(P, A, \DB)$ to $\Cands$
	}

\ForEach{$X \in \Cands$ ordered by the estimate} {
	\If {$L(\DB, \Patterns \cup X) < L(\DB, \Patterns)$} {
		$\Patterns \define \textsc{Prune}(\Patterns \cup X, \DB, \False)$\;
	}
	\lIf {$X$ is added} {
		test recursively $X$ augmented with events occurring in the gaps
	}
}
}

$\Patterns \define \textsc{Prune}(\Patterns, \DB, \True)$\;
order patterns $X \in G$ by $L(\DB, \Patterns) - L(\DB, \Patterns\setminus X)$\;
\Return {$\Patterns$\;}
\end{algorithm}

%% file: related.tex
\section{Related Work}\label{sec:related}

Discovering frequent sequential patterns is an active research topic.
Unlike for itemsets, there are several definitions for frequent
sequential patterns. The first approach counts the number of sequences
containing a pattern~\cite{wang:04:bide}. In such setup, having one long
sequence do not make sense. In the second approach we count multiple occurrences
within a sequence. This can be done by sliding a
window~\cite{mannila:97:discovery} or counting disjoint minimal windows~\cite{laxman:07:fast}.

Mining general episodes, patterns where the order of events are specified by a
DAG is surprisingly hard. For example, testing whether a sequence contains a
pattern is \textbf{NP}-complete~\cite{tatti:11:kdd}. Consequently, research has focused on mining subclasses of episodes, such as, episodes with unique
labels~\cite{achar:11:discovering,pei:06:discovering}, and strict episodes~\cite{tatti:11:clsepdami}.

Discovering statistically significant sequential patterns is a surprisingly
understudied topic. One reason is that unlike for itemsets,
computing an expected frequency under a null-hypothesis is very complex.
Using independence assumption as a null-hypothesis has been suggested
in~\cite{gwadera:05:reliable,tatti:09:significance} and a Markov-chain model has been suggested in~\cite{gwadera:05:markov}.
In~\cite{achar:11:discovering} the authors use information theory-based measure to determine which edges to include in a general episode.

Summarising sequences using segmentation is a well-studied topic. The goal in
segmentation is to divide the sequence in large segments of homogenous regions
whereas our goal is to find a set of compact patterns that occur significantly
often. For an overview in segmentation, see~\cite{dzeroski:10:iqbook}, and for a segmentation tool see~\cite{kiernan:09:eventsum}.

Mannila and Meek~\cite{mannila:00:global} regard general episodes, as generative
models for sequences. Their model generates short sequences by selecting a
subset of events from an episode and select a random order compatible with the
episode. They do not allow gaps and only one pattern is responsible for generating a single sequence. This is not feasible for our setup, where we may have long sequences and many patterns occurring in one sequence.

\SQS draws inspiration from the \textsc{Krimp}~\cite{vreeken:11:krimp} and \textsc{Slim}~\cite{smets:12:slim} algorithms. \textsc{Krimp} pioneered the use of MDL for identifying good pattern sets; specifically, mining sets of itemsets that describe a transaction database well. As serial episodes are much more expressive than itemsets, we here need a much more elaborate encoding scheme, and in particular, a non-trivial approach for covering the data. For mining the patterns, \SQSCands shares the greedy selection over an ordered set of candidates.

Smets and Vreeken recently gave the \textsc{Slim} algorithm~\cite{smets:12:slim} for directly mining \textsc{Krimp} code tables from data. With \SQSSearch we adopt a strategy that resembles \textsc{Slim}, by considering joins $XY$ of $X,Y \in \CT$, and estimating the gain of adding $XY$ to $\CT$. Whereas \textsc{Slim} iteratively searches for the best addition, for efficiency, \SQSSearch adopts a batch-wise strategy.

Lam et al. introduced \textsc{GoKrimp}~\cite{lam:12:gokrimp} for mining sets of serial episodes. As opposed to the MDL principle, they use fixed length codes, and do not punish gaps within patterns---by which their goal is essentially to cover the sequence with as few patterns as possible, which is different from our goal of finding patterns that succinctly summarise the data. 
Bathoorn et al.~\cite{bathoorn:06:freqpatset} also cover greedily, and do not consider gaps at all.

%% file: experiments.tex
\section{Experiments}\label{sec:experiments}
We implemented our algorithms in C++, and provide the source code for research purposes, together with the considered datasets, as well as the generator for the synthetic data.\!\footnote{\url{http://adrem.ua.ac.be/sqs/}}
As candidates for \SQSCands, we mined frequent serial episodes~\cite{tatti:11:clsepdami,laxman:07:fast} using disjoint minimal windows of maximal length $15$, with minimal support thresholds as low as feasible---i.e. at the point where the number of patterns starts to explode. 
All experiments were executed single-threaded on a six-core Intel Xeon machine with 12GB of memory, running Linux.

In our experiments we consider both synthetic and real data. Table~\ref{tbl:results} shows the base statistics per dataset, i.e. number of distinct events, number of sequences, total number of events per database, and the total encoded length by the most simple code table $\ST$.

\emph{Synthetic Data.}
First, we consider the synthetic \textit{Indep},  \textit{Plants-10}, and
\textit{Plants-50} datasets. Each consists of a single sequence of $10\,000$
events over an alphabet of $1\,000$. In the former, all events are independent,
whereas in the latter two we planted resp. $10$ and $50$ patterns of $5$ events
$10$ times each, with $10\%$ probability of having a gap between consecutive events, but
are independent otherwise.

Table~\ref{tbl:results} shows the results given by \SQSCands and \SQSSearch. For the \textit{Indep} dataset, while over $9\,000$ episodes occur at least $2$ times, both methods correctly identify the data does not contain significant structure. 
Similar for \textit{Plants-10} both methods correctly return the $10$ planted patterns. \textit{Plants-50} has a very high density of pattern symbols ($25\%$), and hence poses a harder challenge. \SQSCands and \SQSSearch identify resp. $47$ and $46$ patterns exactly, the remainder consisting of fragments of correct patterns. The imperfections are due to patterns being partly overwritten during the generation of the data. 

\emph{Real Data.}
For the experiments on real data, in order to interpret the patterns, we consider text data. The events are the stemmed words from the text, with stop words removed. \textit{Addresses} contains speeches of American presidents, \textit{JMLR} consists of abstracts of papers from the Journal of Machine Learning Research website,\!\footnote{\url{http://jmlr.csail.mit.edu/}} whereas \textit{Moby} contains the novel `Moby Dick' by Herman Melville.  

Let us first consider the number of returned patterns, as shown in Table~\ref{tbl:results}. We see that for all datasets small numbers of patterns are returned, in the order of $100$s, two orders of magnitude less than the number of frequent patterns \SQSCands considers.

When we consider the gains in compressed size, i.e. $\Delta L = L(\DB,\ST) - L(\DB,\CT)$, we see these few patterns in fact describe a lot of structure of the data; recall that $1$ bit of gain corresponds to an increase of factor $2$ in likelihood. We note \SQSSearch slightly outperforms \SQSCands, which is due to the former being able to consider candidates of lower support without suffering from the pattern explosion.

The largest $\Delta L$ is recorded for \textit{JMLR}, with almost $30$k bits. This is not surprising, as the type of text, abstracts of machine learning papers, has a relatively small vocabulary---the use of which is quite structured, with many key phrases and combinations of words.

Table~\ref{tbl:jmlr} depicts the top-$10$ most compressing patterns for \textit{JMLR}, as found by \SQSSearch. Here, as $\Delta L$ we give the increase in bits the pattern would be removed from $\CT$. Clearly, key machine learning concepts are identified, and importantly, the patterns are neither redundant, nor polluted with common words. In fact, in none of the $\CT$s patterns incorrectly combine frequent events.

\begin{table}
\caption{JMLR data. Top-$10$ patterns by \SQSSearch}\label{tbl:jmlr}
\begin{tabular}{r@{\hspace{0.4em}}ll r@{\hspace{0.4em}}ll}
\toprule
& \textbf{patterns} & $\Delta L$ 
&& \textbf{patterns} & $\Delta L$ \\
\midrule
1. & supp. vector machine & 850 & 
6. & large scale & 329 \\
2. & machine learning & 646 & 
7. & nearest neighbor & 322 \\
3. & state [of the] art & 480 & 
8. & decision tree & 293 \\
4. & data set & 446 & 
9. & neural network & 289  \\
5. & Bayesian network & 374 &
10. & cross validation & 279 \\
\bottomrule
\end{tabular}
\end{table}

Further examples of patterns reported for \textit{JMLR} include `\textit{non neg matrix factor}', `\textit{isotrop log concav distribut}', and `\textit{reproduc[ing] kernel Hilbert space}'. For the presidential \textit{Addresses}, we unsurprisingly see `\textit{unit[ed] stat[es]}' and `\textit{fellow citizen[s]}' as the top-$2$ patterns. An example of a pattern with many gaps ($5.2$ gap events, on average), we find the rather current `\textit{economi[c] public expenditur[e]}'. From the \textit{Moby Dick} novel we find the main antagonist's species, `\textit{sperm whale}', and name, `\textit{moby dick}', as well as the phrase `\textit{seven hundr[ed] seventy seventh}' which occurs $6$ times.

\begin{figure}
\begin{tikzpicture}
\begin{axis}[xlabel=support threshold $\sigma$, ylabel= {$\Delta L$ (bits)},
    width = 4.2cm,
	xmax = 100,
    xtick = {5, 25, 50, 100},
	scaled x ticks = false,
    ytick = {1800, 2500, ..., 5300},
    y tick label style = {/pgf/number format/set thousands separator = {\,}},
    cycle list name=yaf,
    ]

\addplot table[x index = 0, y index = 3, header = false] {results/address_order.dat};

\pgfplotsextra{\yafdrawaxis{5}{100}{1600}{5306}}
\end{axis}
\end{tikzpicture}
\begin{tikzpicture}
\begin{axis}[xlabel = iteration, ylabel= {$\Delta L$ (bits)},
    width = 4.2cm,
	no markers,
	xtick = {5, 35, ..., 155},
    ytick = {1100, 1800, ..., 5300},
	ymax = 5300,
    y tick label style = {/pgf/number format/set thousands separator = {\,}},
    cycle list name=yaf,
	legend pos = south east
    ]

\addplot file {results/address_progress_slam.dat};
\addplot file {results/address_progress_order.dat};
\legend{\textsc{Sqs-Srch}, \textsc{Sqs-Cnd}}

\pgfplotsextra{\yafdrawaxis{1}{155}{900}{5306}}
\end{axis}
\end{tikzpicture}

\caption{\textit{Addresses} dataset, gain in compression. (left) varying support thresholds for \SQSCands. (right) \SQSCands and \SQSSearch per accepted candidate.}
\label{fig:res}
\end{figure}
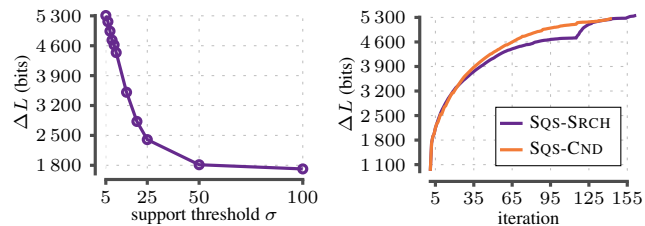

Next, we investigate our search strategies. First, in the left-hand plot of Fig.~\ref{fig:res}, for \SQSCands on the \textit{Moby} data, we show the gain in compression for different support thresholds. It shows that lower thresholds, i.e richer candidate sets, allow for (much) better models---though by the pattern explosion, mining candidates at low $\sigma$ can be infeasible.

Second, in the right-hand plot, we compare \SQSCands and \SQSSearch, showing the gain in bits over $\ST$ per candidate accepted into $\CT$. It shows both search processes are efficient, considering patterns that strongly aid compression first. The slight dip of \SQSSearch around iteration $100$ is due to its batch-wise search. At the expense of extra computation, an iterative search for the best estimated addition, like \textsc{Slim}~\cite{smets:12:slim} may find better models.

In these experiments, using these support thresholds, mining the candidates took up to $4$ minutes, after which \SQSCands took up to $15$ minutes to order and filter the candidates. \SQSSearch resp. took $10$, $18$, and $91$ minutes. As   \textit{Moby} has a large alphabet and is one long sequence, \SQSSearch has to consider many possible pattern co-occurrences.

\begin{table*}[htb!]
\begin{center}
\caption{Basic statistics and results per dataset}\label{tbl:results}
\begin{tabular}{l r rrrrr r rrrrr}
\toprule
&&&&&& \multicolumn{4}{l}{\SQSCands} && \multicolumn{2}{l}{\SQSSearch}\\
\cmidrule{7-10}
\cmidrule{12-13}
\textbf{Dataset} & $\abs{\AB}$ & $\abs{D}$ & $\norm{D}$ & $L(\DB, ST)$ & & $\sigma$ & $\abs{\ifam{F}}$ & $\abs{\ifam{G}}$ & $L(\DB, \CT)$ && $\abs{\ifam{G}}$ & $L(\DB, \CT)$  \\
\midrule
Indep & $1\,000$ & $1$ & $10\,000$ & $103\,630$ &  & $2$ & $9\,094$ & $0$ & $103\,630$ &  & $0$ & $103\,630$\\
Plants-$10$ & $1\,000$ & $1$ & $10\,000$ & $103\,340$ & & $2$ & $11\,957$ & $10$ & $100\,629$ & & $10$ & $100\,629$\\
Plants-$50$ & $1\,000$ & $1$ & $10\,000$ & $102\,630$ & & $2$ & $25\,484$ & $50$ & $91\,706$ & & $52$ & $91\,707$\\
\midrule
Addresses & $5\,295$ & $56$ & $62\,066$ & $685\,593$ &  & $5$ & $15\,506$ & $138$ & $680\,287$ &  & $155$ & $680\,236$\\
JMLR & $3\,846$ & $788$ & $75\,646$ & $772\,112$ &  & $5$ & $40\,879$ & $563$ & $742\,966$ &  & $580$ & $742\,953$\\
Moby & $10\,277$ & $1$ & $105\,719$ & $1\,250\,149$ &  & $5$ & $22\,559$ & $215$ & $1\,240\,667$ &  & $231$ & $1\,240\,566$\\
\bottomrule
\end{tabular}
\end{center}
\end{table*}

%% file: discussion.tex
\section{Discussion}\label{sec:discussion}

The experiments show both \SQSCands and \SQSSearch return high-quality models. By using synthetic data we showed that \SQS reveals the true patterns without redundancy, while further not picking up on spurious structure. Analysis of the results on the text data experiments show that key phrases are identified---combinations of words that may be interspersed with `random' words in the data. Importantly, for all of the datasets, no noisy or redundant patterns are returned. As expected, the more structure a dataset exhibits, the better the attained compression.

With \SQSCands we allow the user the freedom to provide a set of candidate serial episodes. In general, lower thresholds correspond to more candidates,  more candidates correspond to a larger search space, and hence better models. \SQSSearch on the other hand, besides an any-time algorithm, is parameter-free, as it generates and tests  patterns that are estimated to improve the score. 

Both algorithms are fast, our prototype implementations taking only few minutes on the data here considered. The algorithms have many opportunities for parallelisation: candidates can be estimated or evaluated individually, as can the scanning for minimal windows. 

MDL does not provide a free lunch. First of all, although highly desirable, it is not trivial to bound the score. While for Kolmogorov complexity we know this is incomputable, for our models we have no proof one way or another. Furthermore, although MDL gives a principled way to construct an encoding, this involves many choices that determine what structure is rewarded. As such, we do not claim our encoding is suited for all goals, nor that it cannot be improved.

Future work includes the extension of \SQS for parallel and general episodes---which surpass serial episodes in expressiveness. Although seemingly opposed to MDL (why describe the same thing twice?) allowing patterns to overlap  may provide more succinct summarisations. Last, but not least, we are interested in applying the \SQS code tables for clustering and anomaly detection.

%% file: conclusion.tex
\section{Conclusion}\label{sec:conclusion}

In this paper we employed the MDL principle to mine sets of sequential patterns that summarise the data well. In particular, we formalised how to encode sequential data using set of patterns, and use the encoded length as a quality measure. 
As search strategy for good models, we adopt two approaches.
The first algorithm, \SQSCands, selects a good pattern set from a large candidate set, while \SQSSearch is a parameter-free any-time algorithm that discovers good pattern sets directly from the data. 
Experimentation on synthetic and real data showed both methods to efficiently discover small, non-redundant sets of informative patterns. 

And that's the long and the short of it.

%% file: appendix.tex
\appendix
\section{Proofs} 
\label{sec:apx}

\begin{proof}[of Proposition~\ref{prop:minimal}]
Assume opposite: there is a window $(i, j, X, k) \in A$ such that $W = S_k[i, j]$ is
not an minimal window for $X$. Let $S_k[a, b]$ be a minimal window of $X$ in $W$.
Let $A'$ be an alignment in which we replace $(i, j, X, k)$ with $(a, b, X, k)$. Note
that $\usage(Y)$ remains constant for any pattern $Y$. In addition, $\gaps(Y)$
remains also constant for any pattern $Y \neq X$. Since $b - a < j - i$, we see
that $\gaps(X) < \gaps(Y)$. A straightforward computation shows that $L(\CT, D)$
is a monotonic function of $\gaps(X)$. Hence, the encoding length of 
$A'$ is lower than of $A$, which contradicts the optimality of $A$.
\end{proof}

\begin{proof}[of Proposition~\ref{prop:align}]
Let

\[
\begin{split}
	& \mathit{const} = \\
	&\quad  L_\mathbb{N}(|\DB|) + \sum_{S \in \DB}{L_\mathbb{N}(|S|)} + \sum_{s \in \Sigma} \supp(s \mid \DB)L(\code_p(s)) \quad.
\end{split}
\]
The first term in the definition of  $\gain$ will introduce the correct number
of usages of non-singleton patterns. The second and the third terms correspond
to the length of the gap stream. Finally, since for $s \in \AB$,
\[
\begin{split}
	\usage(s) & = \supp(s \mid \DB) - \sum_{s \in X} \usage(X) \\
	          & = \supp(s \mid \DB) - \sum_{s \in X} \abs{\set{(i, j, X, k) \in A}}\quad,
\end{split}
\]
the fourth term will correctly reduce the singleton usages. 
\end{proof}

\begin{proof}[of Proposition~\ref{prop:gainconstant}]
We will assume that $P \neq Q$, the treatment for the case $P = Q$ is almost
equivalent.  Let $A' = A \cup A \setminus (V \cup W)$. The usages in $A'$
remain the same except for $P$, $Q$, and $R$: Usages for $P$ and $Q$ are
reduced by $N$ and usage of $R$ is increased by $N$. In addition, the total
usage $\usage(\CT(A')) = \usage(\CT(A)) - N$ is reduced by $N$.

To compute the difference in the pattern stream $C_p$ we first compute the
difference between the code lengths for patterns $P$, $Q$, and $R$ using new usages
but old total usage. We have $\usage(\CT(A'))$ with incorrect total usage.
To compensate the difference in total usage we add
\[
	\usage(\CT(A'))(\log \usage(\CT(A')) - \log \usage(\CT(A)))\quad.
\]

The gaps $\gaps(P)$ and $\gaps(Q)$ are decreased by $\gaps(V)$ and $\gaps(W)$
under the new encoding. Also, $\gaps(R)$ is increased by $\gaps(U)$. The
remaining gaps remain the same. Consequently we can compute the difference in
the gap stream $C_g$ in constant time.
Hence, we can compute the difference $L(D \mid A') - L(D \mid A)$ in constant time.

Encoding the code table will change since it depends on total usage of
non-singleton patterns.  In addition, we may delete $P$ or $Q$ from the code
table if their usage counts go to zero (or add $R$ it its usage count was $0$).
We see from the definition of $L(CT)$ that in total $6$ terms may change.
Consequently, we can compute the difference $L(CT(A')) - L(CT(A))$ in constant
time.
\end{proof}

\begin{proof}[of Proposition~\ref{prop:smallest}]
Let $A' = A \cup \set{w} \setminus \set{v}$. The usage counts in $A$ and in
$A'$ are the same. Thus $L(C_p \mid \CT(A')) = L(C_p \mid CT(A))$.
The gaps also remain constant except for $X$, in which case,
the $\gaps(X)$ is reduced by $i - j - (b - a)$. A straightforward calculation
implies that $L(C_p \mid \CT(A')) < L(C_p \mid \CT(A))$ and $L(X, \CT(A')) < L(X, \CT(A))$.
This implies that $L(D, A') < L(D, A)$.
\end{proof}

\begin{proof}[of Proposition~\ref{prop:optimal}]
Let $\mathit{PX}$ be the optimal pattern.  Since alignment is empty, we do not
need to compensate for overlapping windows and the encoding lengths we are
computing are accurate.  The algorithm enumerates windows from smallest to
largest.  We can use Proposition~\ref{prop:smallest} to see that there will be
a point where $U_X$ will contain the optimal alignment, yielding a correct optimal $d_X$. 
Consequently, \textsc{Estimate} will return $\mathit{PX}$.
\end{proof}

%% file: paper.bbl
\providecommand{\noopsort}[1]{}
\begin{thebibliography}{10}

\bibitem{achar:11:discovering}
A.~Achar, S.~Laxman, R.~Viswanathan, and P.~S. Sastry.
\newblock Discovering injective episodes with general partial orders.
\newblock {\em Data Min.\ Knowl.\ Disc.}, 2011.

\bibitem{bathoorn:06:freqpatset}
R.~Bathoorn, A.~Koopman, and A.~Siebes.
\newblock Reducing the frequent pattern set.
\newblock In {\em ICDM-Workshop}, pages 1--5, 2006.

\bibitem{cover:06:elements}
T.~M. Cover and J.~A. Thomas.
\newblock {\em Elements of Information Theory}.
\newblock Wiley-Interscience New York, 2006.

\bibitem{dzeroski:10:iqbook}
S.~Dzeroski, B.~Goethals, and P.~Panov, editors.
\newblock {\em Inductive Databases and Constraint-Based Data Mining}.
\newblock Springer, 2010.

\bibitem{grunwald:07:book}
P.~Gr\"{u}nwald.
\newblock {\em The Minimum Description Length Principle}.
\newblock MIT Press, 2007.

\bibitem{gwadera:05:markov}
R.~Gwadera, M.~J. Atallah, and W.~Szpankowski.
\newblock Markov models for identification of significant episodes.
\newblock In {\em SDM}, pages 404--414, 2005.

\bibitem{gwadera:05:reliable}
R.~Gwadera, M.~J. Atallah, and W.~Szpankowski.
\newblock Reliable detection of episodes in event sequences.
\newblock {\em Knowl. Inf. Sys.}, 7(4):415--437, 2005.

\bibitem{kiernan:09:eventsum}
J.~Kiernan and E.~Terzi.
\newblock {EventSummarizer}: a tool for summarizing large event sequences.
\newblock In {\em EDBT}, pages 1136--1139, 2009.

\bibitem{lam:12:gokrimp}
H.~T. Lam, F.~M\"{o}rchen, D.~Fradkin, and T.~Calders.
\newblock Mining compressing sequential patterns.
\newblock In {\em SDM}, 2012.

\bibitem{laxman:07:fast}
S.~Laxman, P.~S. Sastry, and K.~P. Unnikrishnan.
\newblock A fast algorithm for finding frequent episodes in event streams.
\newblock In {\em KDD}, pages 410--419, 2007.

\bibitem{vitanyi:93:book}
M.~Li and P.~Vit\'{a}nyi.
\newblock {\em An Introduction to Kolmogorov Complexity and its Applications}.
\newblock Springer, 1993.

\bibitem{mannila:00:global}
H.~Mannila and C.~Meek.
\newblock Global partial orders from sequential data.
\newblock In {\em KDD}, pages 161--168, 2000.

\bibitem{mannila:97:discovery}
H.~Mannila, H.~Toivonen, and A.~I. Verkamo.
\newblock Discovery of frequent episodes in event sequences.
\newblock {\em Data Min.\ Knowl.\ Disc.}, 1(3):259--289, 1997.

\bibitem{pei:06:discovering}
J.~Pei, H.~Wang, J.~Liu, K.~Wang, J.~Wang, and P.~S. Yu.
\newblock Discovering frequent closed partial orders from strings.
\newblock {\em IEEE TKDE}, 18(11):1467--1481, 2006.

\bibitem{rissanen:83:integers}
J.~Rissanen.
\newblock Modeling by shortest data description.
\newblock {\em Annals Stat.}, 11(2):416--431, 1983.

\bibitem{salomon:09:handbook}
D.~Salomon and G.~Motta.
\newblock {\em Handbook of Data Compression}.
\newblock Springer, 2009.

\bibitem{smets:12:slim}
K.~Smets and J.~Vreeken.
\newblock \textsc{Slim}: Directly mining descriptive patterns.
\newblock In {\em SDM}, pages 1--12. SIAM, 2012.

\bibitem{tatti:09:significance}
N.~Tatti.
\newblock Significance of episodes based on minimal windows.
\newblock In {\em ICDM}, pages 513--522, 2009.

\bibitem{tatti:11:kdd}
N.~Tatti and B.~Cule.
\newblock Mining closed episodes with simultaneous events.
\newblock In {\em KDD}, pages 1172--1180, 2011.

\bibitem{tatti:11:clsepdami}
N.~Tatti and B.~Cule.
\newblock Mining closed strict episodes.
\newblock {\em Data Min.\ Knowl.\ Disc.}, 2011.

\bibitem{vereshchagin:03:kolmo}
N.~Vereshchagin and P.~Vitanyi.
\newblock Kolmogorov's structure functions and model selection.
\newblock {\em IEEE TIT}, 50(12):3265-- 3290, 2004.

\bibitem{vreeken:08:misval}
J.~Vreeken and A.~Siebes.
\newblock Filling in the blanks: Krimp minimisation for missing data.
\newblock In {\em ICDM}, pages 1067--1072, 2008.

\bibitem{vreeken:11:krimp}
J.~Vreeken, M.~{van Leeuwen}, and A.~Siebes.
\newblock \textsc{Krimp}: Mining itemsets that compress.
\newblock {\em Data Min.\ Knowl.\ Disc.}, 23(1):169--214, 2011.

\bibitem{wang:04:bide}
J.~Wang and J.~Han.
\newblock Bide: Efficient mining of frequent closed sequences.
\newblock {\em ICDE}, 0:79, 2004.

\end{thebibliography}
